\documentclass[twocolumn,floatfix,superscriptaddress,nofootinbib,longbibliography,aps]{revtex4-2}
\usepackage{amsmath,amsfonts,amssymb}
\usepackage{bm}
\usepackage{physics}
\usepackage{xcolor}
\usepackage[pdftex]{graphicx}
\usepackage{float}
\usepackage{changepage}
\usepackage{tabularx}
\usepackage{setspace}
\usepackage[font=small, labelfont=bf, justification=raggedright]{caption}

\usepackage{geometry}

\usepackage[font=small,labelfont=bf]{caption}

\newcommand{\heading}[1]{{\vspace{0.25truecm}\noindent\textbf{#1.}}}

\begin{document}

\title{The voice of few, the opinions of many: evidence of social biases in Twitter COVID-19 fake news sharing}

\author{Piergiorgio Castioni}
\affiliation{Istituto di Scienze e Tecnologie della Cognizione }
\affiliation{Departament d’Enginyeria Informàtica i Matemàtiques, Universitat Rovira i Virgili, 43007 Tarragona, Spain}

\author{Giulia Andrighetto}
\affiliation{Istituto di Scienze e Tecnologie della Cognizione }

\author{Riccardo Gallotti}
\affiliation{CoMuNe Lab, Fondazione Bruno Kessler, Via Sommarive 18, 38123 Povo (TN), Italy}

\author{Eugenia Polizzi}
\affiliation{Istituto di Scienze e Tecnologie della Cognizione }

\author{Manlio De Domenico}
\affiliation{CoMuNe Lab, Fondazione Bruno Kessler, Via Sommarive 18, 38123 Povo (TN), Italy}
\affiliation{Department of Physics and Astronomy ``Galileo Galilei'', University of Padova, Padova, Italy}

\begin{abstract}
Online platforms play a relevant role in the creation and diffusion of false or misleading news. Concerningly, the COVID-19 pandemic is shaping a communication network -- barely considered in the literature -- which reflects the emergence of collective attention towards a topic that rapidly gained universal interest. Here, we characterize the dynamics of this network on Twitter, analyzing how unreliable content distributes among its users. We find that a minority of accounts is responsible for the majority of the misinformation circulating online, and identify two categories of users: a few active ones, playing the role of ``creators'', and a majority playing the role of ``consumers". The relative proportion of these groups ($\approx$ 14\% creators -- 86\% consumers) appears stable over time: Consumers are mostly exposed to the opinions of a vocal minority of creators, that could be mistakenly understood as of representative of the majority of users. The corresponding pressure from a perceived majority is identified as a potential driver of the ongoing COVID-19 infodemic.
\end{abstract}

\date{\today}

\maketitle

\section*{Introduction}

The spread of COVID-19, a respiratory disease responsible for the emerging pandemic observed in early 2020 to date~\cite{hoehl2020evidence,kraemer2020effect,aleta2020modelling}, has led to an increase in misinformation and disinformation about a broad range of topics, from health to technology~\cite{roozenbeek2020susceptibility}. However, due to the unprecedented global health crisis we are facing, the rise of fake news has become a global concern, with the potential to affect public health policy and public order~\cite{rapp_cant_2018,earnshaw2020covid}. The World Health Organization has recognized this phenomenon and has referred to as ``infodemic''~\cite{who2020}, a massive volume of news and narratives not necessarily reliable, including a variety of false rumors and unreliable news appearing during a disease outbreak \cite{organization_managing_2018}, identifying in artificial intelligence an invaluable tool to support the global response against it~\cite{luengo2020artificial}. In fact, coupling misinformation spreading with an ongoing pandemic might be particularly dangerous for public health~\cite{tangcharoensathien2020framework,calleja2021public}, as it erodes trust institutions~\cite{pummerer2020conspiracy} and can lead people to turn to ineffective -- and potentially harmful -- remedies, as well as to engage in risky behavior for them and for the collectivity (e.g., refusing vaccines or not adopting non-pharmaceutical interventions, such as wearing masks and physical distancing)~\cite{van2020using} that substantially increase epidemic spread~\cite{organization_managing_2018,waszak_spread_2018,gallotti_assessing_2020}. \\ 
Over the past years, social media platforms have become the preferred arena for public debates. Notwithstanding, they also represent some of the most vulnerable hotspots for infodemics to occur~\cite{frenkel_surge_2020,russonello_afraid_2020,loomba2021measuring}. For instance, on social media platforms, false information is typically shared by more users, and travels far more rapidly than reliable information \cite{vosoughi_spread_2018}. This condition is further exacerbated by the presence of social bots~\cite{Ferrara2016,cresci2021coming} -- i.e., automated accounts impersonating humans -- that act as magnifiers of noise, conflicts and (mis)information spread~\cite{shao2018spread,Stella2018,ross_are_2019,stella2019influence,chen2021neutral,gonzalez2021bots,caldarelli2021flow}. Fed by users' preferences and attitudes, social media algorithms increase the selectivity of the contents to which users are exposed, further limiting content verifiability \cite{ciampaglia_how_2018}. All this has been shown to provide a fertile ground for the emergence of echo chambers and epistemic bubbles \cite{nguyen_2020}, well-formed and highly segregated communities where unsubstantiated rumors can gain increased exposure and become highly resistant to correction \cite{Garrett2013ThePA}. \\

\begin{figure*}[t]
    \centering
    \includegraphics[width=\textwidth]{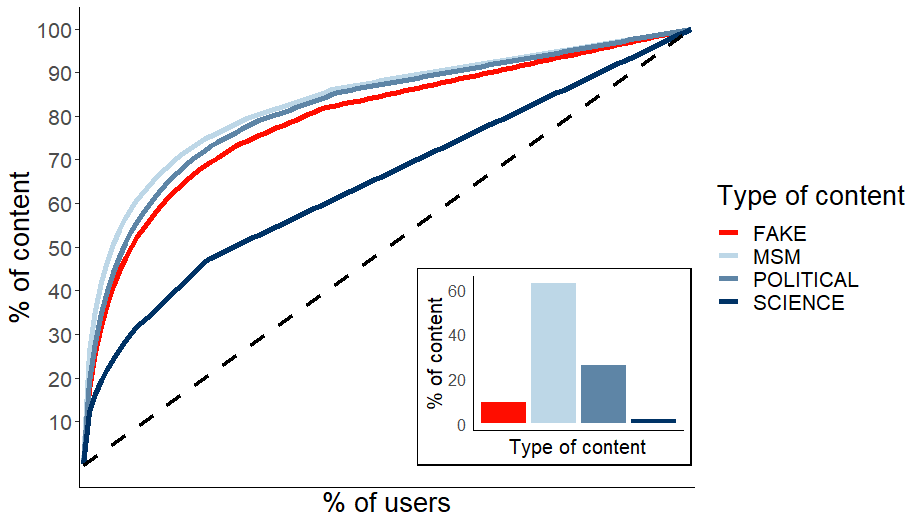}
    \caption{\textbf{Characterizing the share of contents with respect to the share of users who produce them.} The $x$-axis indicates the share of users appearing more frequently in our data set, while the $y$-axis displays the share of the overall content (tweets and retweets) that those users are responsible for. Different content types are encoded by distinct colours: note the red one, corresponding to content identified as fake news. Black dashed lines correspond to the distribution one would observe if all users were responsible for the same fraction of total content, to highlight the highly heterogeneous activity of content production from different users, regardless of content type.
    }
    \label{fig:quantile}
\end{figure*}

In this paper, we investigate factors that may facilitate the share of misinformation in online settings, and we focus our attention on the role of social influence. 
Social psychology has shown that beliefs and actions of individuals are heavily affected by what they perceive the majority of their peers thinks or does \cite{Cialdini1999}. Yet this perception does not need to be necessarily correct to be effective. A wrong perception of what the majority opinions and actions leads to what in social psychology is known as “pluralistic ignorance'' \cite{katz_students_1931}, a situation in which the majority of members of a group privately disagree with a certain opinion or social norm, yet (incorrectly) believe that the other group members accept it. Under this situation, individuals believe that they will lose social reputation or will receive negative reactions from other members of their group if they behave how they wish. Therefore, people are likely to act in accordance with the perceived majority even if they privately disapprove of the behavior. Consequently, public behaviors of groups as a whole do not coincide with the majority of group members' private preferences under circumstances of pluralistic ignorance. Previous research has examined the role of pluralistic ignorance across many topics, including heavy drinking \cite{prentice_pluralistic_1993}, racial segregation \cite{fields_public_1976}, casual sex \cite{lambert_pluralistic_2003}, bullying \cite{sandstrom_social_2013}, adolescent delinquency \cite{young_delinquency_2013}, stigmatization at work \cite{munsch_pluralistic_2014}, providing evidence for its effect in supporting the persistence of unpopular attitudes and behavior.
Despite its potential role in favouring the spread of misinformation, evidences of pluralistic ignorance in social media require still to be directly addressed. We thus contribute to this line of research by investigating whether online communication network features -- e.g., those based on content sharing-- play a role in developing a misperception about the opinion of the majority. Specifically, we suggest that a mismatch -- e.g., both in terms of size and in terms of segmentation -- between users responsible for producing unreliable contents and those that mainly share unreliable contents may represent a structural basis on which pluralistic ignorance can naturally emerge.
To this aim, we provide a quantitative analysis of empirical human activities gathered from Twitter, a popular micro blogging platform, on unreliable contents --such as fake news and conspiracy theories -- in the context of COVID-19. 
For this purpose we collected and analysed 7.7 million retweets belonging to 1.6 million users and we introduce a criterion for characterizing the users responsible for spreading fake news by means of two groups that we name 'creators' and 'consumers'. 
Our choice is supported by empirical evidence provided in first section of this work. We show that the size of the two groups differs significantly: the creators are almost 15k while the consumers are 93k, amounting to the 14\% and 86\% of the fake news spreaders population, respectively.
In the second section we prove that such a definition is solid and suitable even when the underlying socio-technical system is analyzed from a dynamic perspective, showing that users tend to mostly remain in the same group and that inter-group switches are mostly temporary. Finally, we analyse the causality relation between the overall volume of fake news and the size of the groups of creators and consumers, finding the latter to be, in fact, a good control parameter to describe the behavior of spread dynamics of unreliable content.

\begin{figure*}[t]
    \centering
    \includegraphics[width=0.8\textwidth]{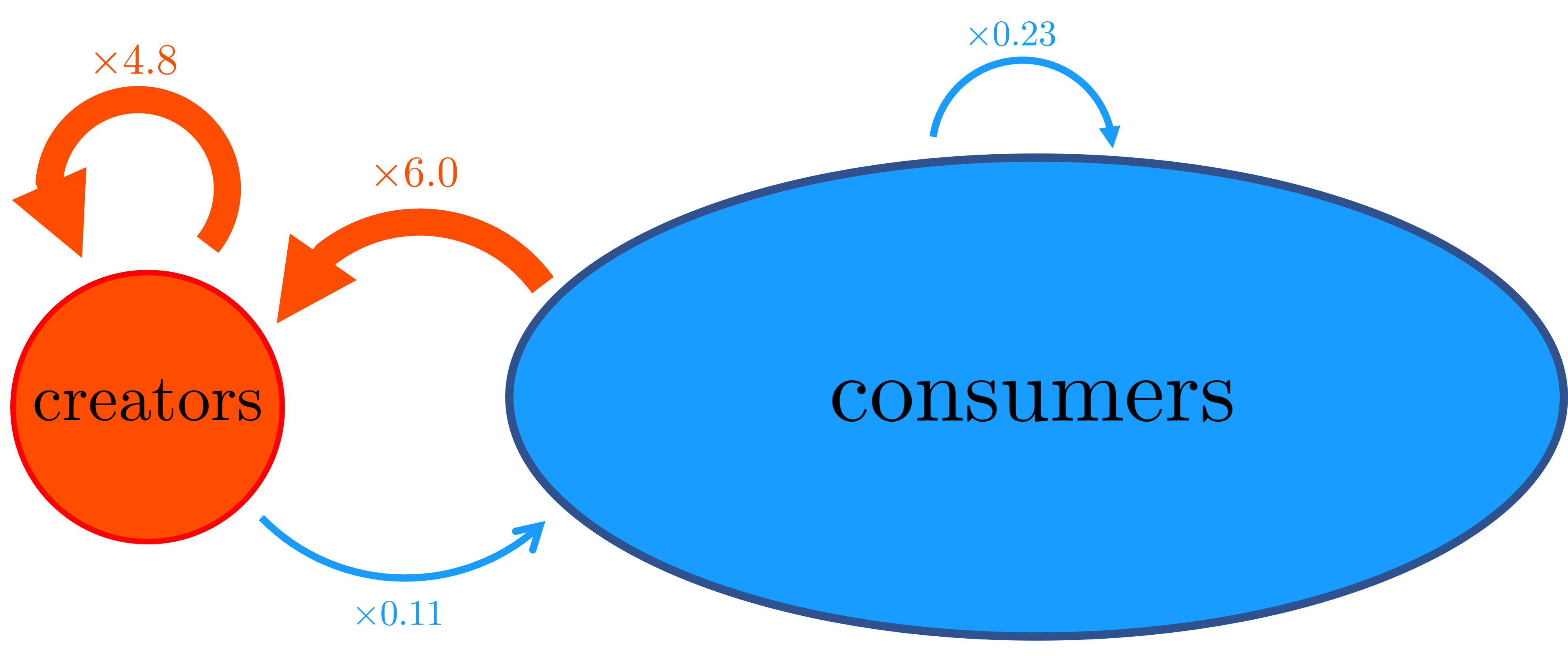}
    \caption{\textbf{Schematic illustration of the separation between creators and consumers}. The arrows represent the endorsements (i.e., retweets in Twitter) and go from the retweeting to the retweeted individuals. Values indicates the ratio between the observed number of links and the number one would expect if the links were randomly assigned. Note that the illustration is not at scale with numbers. }
    \label{fig:schema}
\end{figure*}

\section*{Results}

\heading{Characterizing the separation between creators and consumers} We analyzed the interaction networks involving users posting messages related to the COVID-19 on a period of time spanning from January the 22nd to May the 22nd 2020. We focused our attention on the U.S., which is a geographic area particularly active on Twitter, where we have passively collected 7.7 million retweets that allowed us to reconstruct the communication flow network between 1.6 million user accounts. 

Users' social behavior, measured in terms of volumes of activities, is not uniformly distributed: instead, it exhibits a heavy-tail distribution, where a minority of individuals is responsible for the great majority of content, as shown in Fig.~\ref{fig:quantile}. Furthermore, this result is confirmed even after stratifying for the type of content: In fact, a minority of users is actively involved in the production of false or misleading content in the context of COVID-19, contributing in a major way to the overall volume of fake news and to the possible over-representation of ideas that would be otherwise be minor. Other users, instead, are characterized by a 'bot-like' behaviour, i.e., the tendency to passively retweet something without commenting on it or appending any original content, thus contributing nonetheless to its diffusion. Users within the platform may assume such misleading information as being representative of what the majority of users believes, which is the basis on which pluralistic ignorance can naturally emerge. Although these two qualitatively different types of users are both ``fake news spreaders'' -- which is the word that we will use to identify them throughout this work -- it is safe to assume that they will have different roles in the misinformation ecosystem.
Users belonging to the first, active, group are highly motivated users, prepared to spend energy and time to craft messages (i.e., tweets) to spread them further. We call a user in this first group as a ``creator''. The second, passive, group contains those users who are more likely to engage in low-cost behaviours, such as retweeting. We will refer to a user in this group as a ``consumer''.The latter user is the one that we suggest may be vulnerable to pluralistic ignorance, namely acting in accordance with the perceived majority, even if he/she privately disapprove of the behavior. Finally, outside of these two groups remains the rest of the online population, consisting of those user accounts that have never shared unreliable content but that, nonetheless, interact with creator and consumer through other kinds of retweets. \\ 

In order to talk more quantitatively about these two groups, it is useful to define them in terms of the fraction of fake retweets per user in a given time window. Specifically, we defined as being creators all those users whose retweeted content in a given time interval, at least 20\% fake. For more details about this choice see Methods. Furthermore, see SI for an analysis of the difference in number of followers between creators and consumers that further supports our intuition behind the definition of the two groups. 


\heading{Fluid transitions between creator and consumer groups} One of the consequences of defining both creators and consumers in terms of user activity is that when the underlying behaviour changes, the composition of the corresponding groups changes accordingly. Operationally, this dynamical behavior over time leads to users going from being active spreaders to being silent, or viceversa, as well as anything in between. In practice, the groups of creators and consumers are fluid constructs that exist at every time step but are also continuously mutating and experiencing inward and outward user flows. \\
It is natural to wonder if the classification provided in the previous section -- that of creators and consumers of fake news -- still holds when such concepts are defined taking short time steps, e.g. one day, while allowing users to switch between groups. We find that only a small minority -- specifically, the 4.67\% -- of all fake news spreaders go from being creators to being consumers or viceversa. Moreover the majority of fake news spreader tends to keep a similar spreading behavior, even after a long period of inactivity (see Table \ref{tab:table}).

\begin{figure*}[t]
    \centering
    \includegraphics[width=\textwidth]{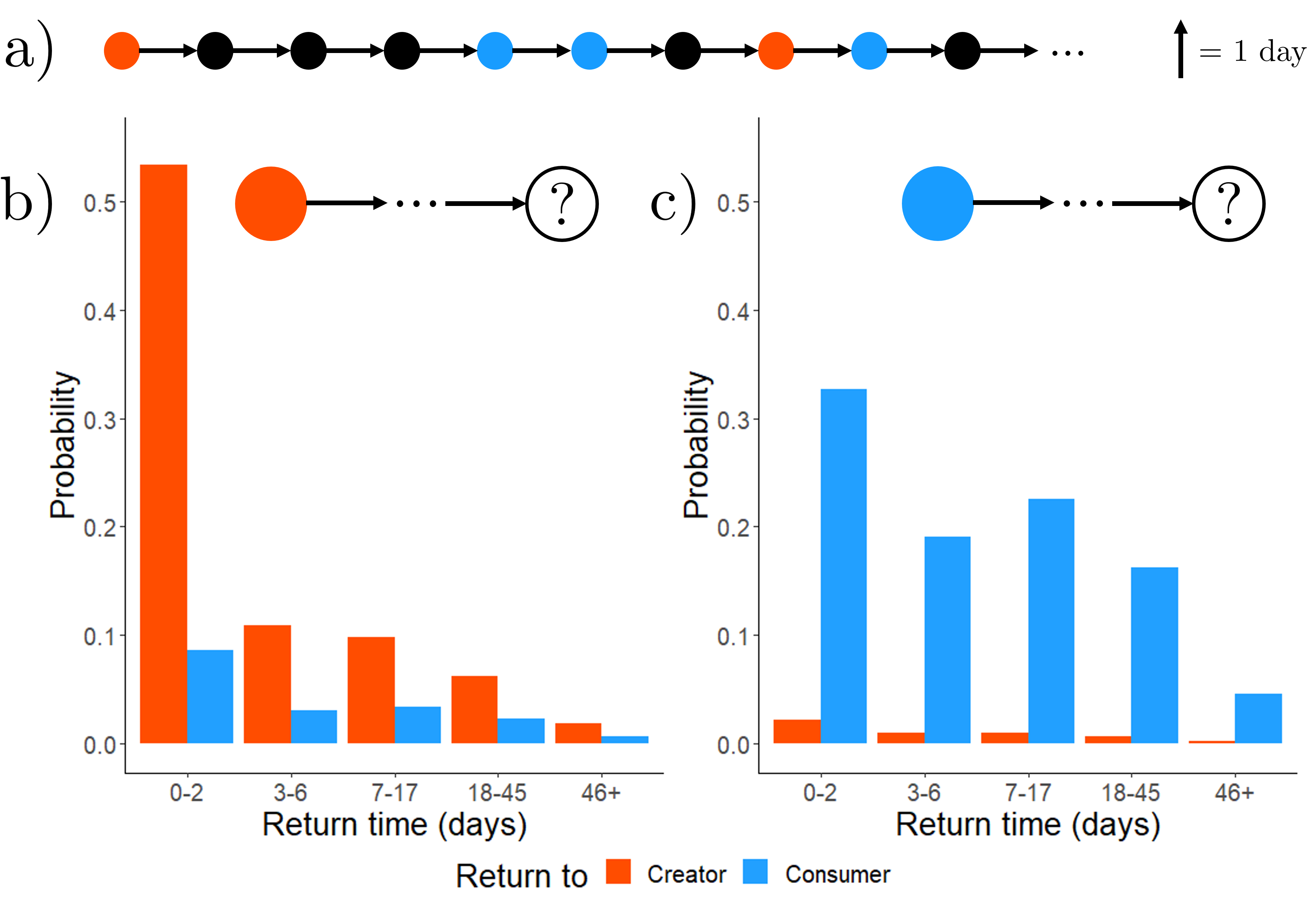}
    \caption{\textbf{Fluid transitions between creator and consumer groups.} Let us consider first-return times: $a)$ schematic example of the behaviour of a user (the circle), who might change his/her group at every time step (e.g., 1 day). Red, blue and black circles represent creators, consumers and non-spreaders, respectively. The return times are the number of black circles that separate colored circles from one another, so in this example they are 3, 0, 1 and 0 days, in chronological order from left-hand to right-hand side. We also report the probability of returning to a skeptic (i.e., non-spreader) group for a user that has just left the creators $b)$ or the consumers $c)$. }
    \label{fig:flow}
\end{figure*}

Fig.~\ref{fig:flow} shows how a user belonging to the creators (or consumers) group is more likely to behave next, and after how much time. The analysis is restricted only to those users for which such a time can be unambiguously identified, meaning that we only consider those users that return to one of the two fake news groups within the time span of our temporal data. Our analysis shows two facts: first, the probability of returning to either one of the two fake news groups decreases with time; second, it is always more likely to return to the same group that a user left than in the other one. In practice, this means that creators are more likely to go back being creators, no matter how much time they spent being silent, and the same hold for consumers. While these features are the same whether we consider creators or consumers, these two groups also display some notable differences. For instance users leaving the creator group are more likely to come back to it within 0--2 days, although in the same time span they can change group with 10\% probability. Conversely, if a user starts from the consumer group it is much more likely to go back to it even after quite long time, while the probability to change group in the same period is much lower, going down to 24 times less likely in the 18-45 days range. This makes this group more stable in terms of user dynamics and, consequently, more suitable to be considered for potential interventions. 
This finding supports our interpretation of consumers as those users who are not very committed to fake news spreading, since the inter-time between consecutive fake news sharing is longer, on average, than that measured for the creators. 

\begin{table}[ht]
    \centering
    \small
    \begin{tabular}{ l|c|c }
        \textbf{Behaviour} & \textbf{Users} & \textbf{Only once}  \\
        \hline
        Only creators & 13848 (12.91\%) & 9804 (9.14\%) \\
        Only consumers & 88400 (82.42\%) & 58328 (54.38\%) \\
        Mixed & 5005 (4.67\%) & {/} 
    \end{tabular}
    \caption{\textbf{User behavior in spreading fake news.} Number of users classified according to their behaviour towards the spread of unreliable content: the majority of them tends to spread fake news only once and, consequently, they were not taken into account in the analysis shown in Fig. \ref{fig:flow}. }
    
    \label{tab:table}
\end{table}

\begin{figure*}[ht]
    \centering
    \includegraphics[width = \textwidth]{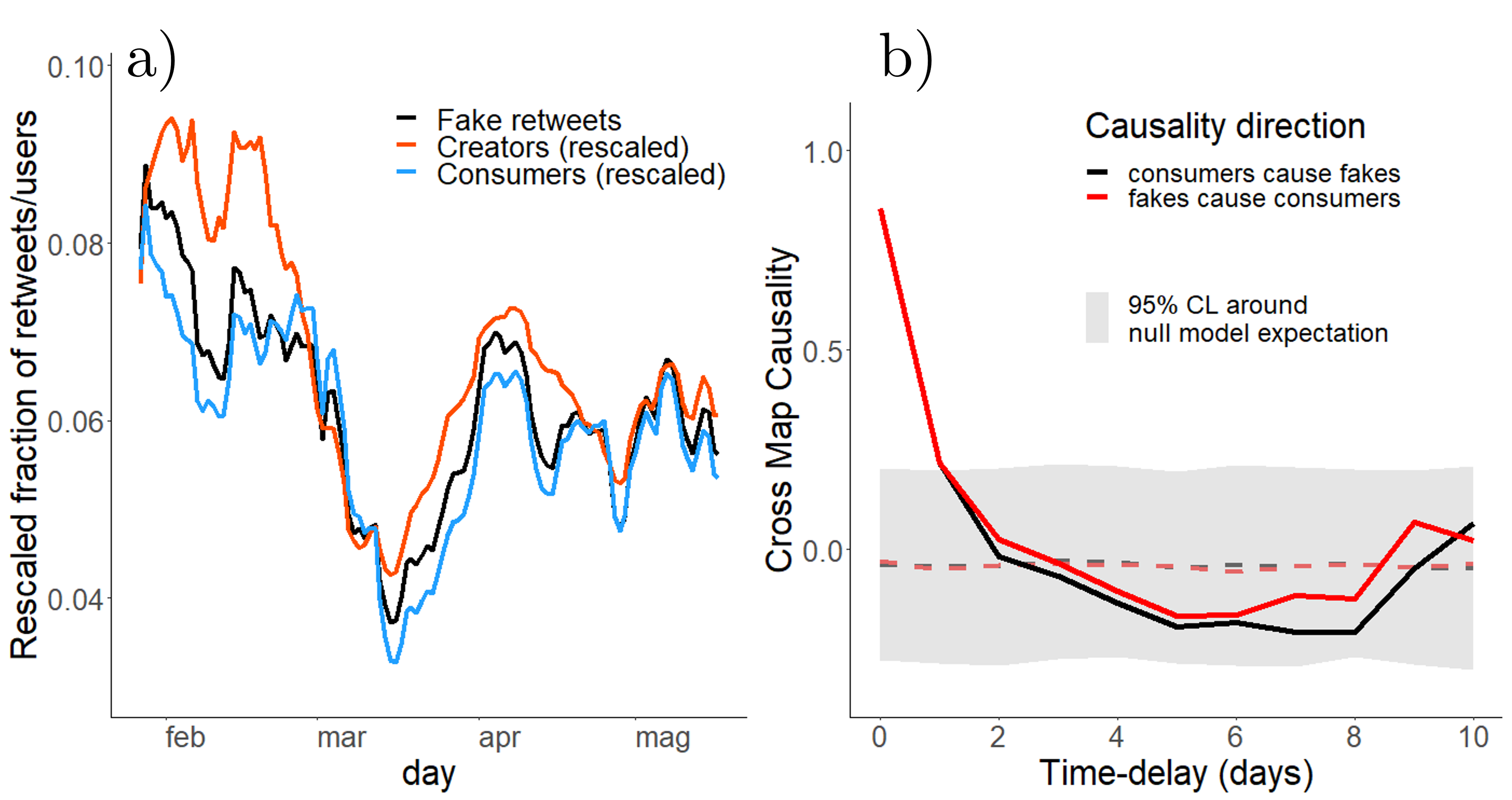}
    \caption{\textbf{Unraveling causal relationships between group dynamics and fake news volume.}
        $a)$ Comparison between time series of the fraction of fake retweets (black line), the fraction of consumers (blue line) and the fraction of creators (red line). The latters were rescaled to ease the comparison between trends. The time step is of one day while the lines are obtained through a ten days moving average.
        $b)$ Cross map signal computed for different time-delays with the Convergent Cross Mapping algorithm (see Methods). For the null hypothesis we used surrogates obtained by randomly reshuffling empirical observations. The null hypothesis is rejected at 95\% CL, equivalent to an \emph{a priori} test size of 5\%, only at time delay equal to 0 and 1 day, with the strongest signal at the former.
     }
    \label{fig:causation}
\end{figure*}

\heading{Unraveling causality between creators' dynamics and fake news volume} It is plausible to ask whether the overall volume of fake news increases in response to an increase in the number of fake news spreaders or if it is the other way around. To inspect the existence of a causal relation between the size of the two communities of fake news spreaders and the overall volume of fake news circulating in the network, we consider temporal snapshots of 1 day for the analysis. For each time slice, we count how many creators and consumers there are and how many fake news are being shared, building the three time series showed in Fig.~\ref{fig:causation}a which appear to be at least correlated with each other. First, we quantify such correlations: The cross-correlation between the fraction of creators and the fraction of fake content is 0.7; this value goes to 0.69 when we consider the fraction of creators and consumers and reaches 0.97 when we consider the last pair of time series, i.e., fraction consumers and fraction of fake content. \\
Moving beyond simple correlations, we identify what is the cause-effect relation between the number of consumers and fake news spreading by means of the Convergent Cross Mapping (CCM) algorithm (see Methods). To investigate all the possible cases we look for both short-time and long-time effects by varying the time-delay parameter that the method allows one to tune. Our results are shown in Fig.~\ref{fig:causation}b, where the null hypothesis is obtained by performing the same causal test over different surrogates of our original time-series, obtained by randomly permuting empirical observations while destroying any temporal correlation.  \\
As shown in Fig. \ref{fig:causation}b, the cross map causality, i.e. the quantity that indicates how strongly two time series are causally linked, is the highest when the time delay is 0 days, and above the 95\% CL only for 0 and 1 days for both the causality directions (consumers cause a higher volume of fake news and such a volume, in turn, cause a growth in consumers size). An analogous result is obtained when, instead of the consumers' dynamics, creators' dynamics is considered. Although a causal relation between the time course of the two empirical dynamics has been found, it is not possible to conclude which of the two quantities is responsible for the variation of the other, either because 1) there is a very strong feedback loop that propagates causal effects over fast time scales; or 2) the two variables are driven by a third, external or hidden, common cause. In either case this proves that the creators-consumers separation that we introduced to explain the fake news spreading is such a fundamental features that it is practically indistinguishable from the fake news themselves.

\begin{figure*}[t]
    \centering
    \includegraphics[width=\textwidth]{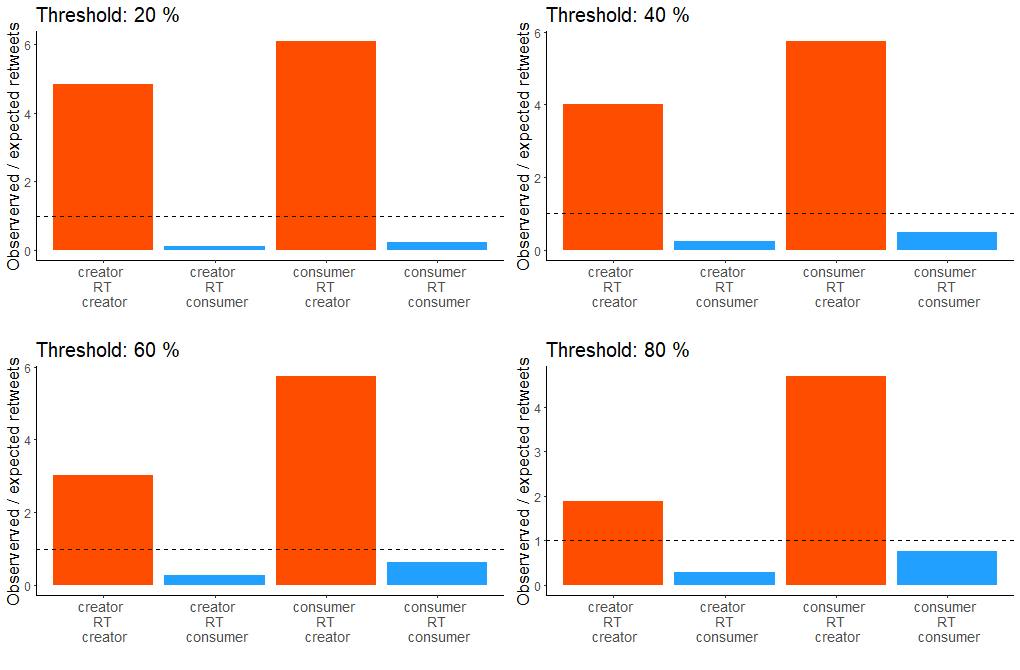}
    \caption{\textbf{Analysis of the separation between creators and consumers for different definitions of these groups.} The height of the bars indicate the ratio between the observed number of links between two groups and the number we would expect if the links were randomly distributed among the network. The dashed horizontal line corresponds to the case where the number of observed links is compatible with those of a random network ($y=1$). The red and blue colors indicate if the tweet was originally from the creators or from the consumers, respectively. The figures differ because of the threshold in the percentage of most active fake news spreaders used to define creators and consumers. However, it can be seen that the densities of the connections between these groups do not depend strongly on such a threshold.  }
    \label{fig:density}
\end{figure*}

\section*{Discussion}

Overall, from our analysis, a consistent picture arises. First of all, we were able to separate fake news spreaders in two groups: a small one (14\%) of active and motivated users (creators) and larger one (86\%) of users who instead prefer to repeat what other people say without creating content on their own (consumers). We did so by putting together an information coming from the structural network such as the in-degree of each node with the information coming from the metadata on the retweets nature, namely the subject of the message. Furthermore the criterion that we chose proved to be robust against the variation of the only arbitrary parameter that it takes as input. \\
As a result we were able to classify all users in the network based on their group's belonging. This allows us to quantify their role in the spreading of fake news and to understand how much they are susceptible to developing a wrong representation of the opinions and beliefs of their community. In particular, the creators strongly interact among themselves, so it is reasonable to think that they have a correct estimate of the opinions of the other members of their group. 
By contrast, the consumers are users who read and retweet contents coming from the creators. Since creators are only a minor fraction of the twitter landscape, such an over exposure to their tweets may lead consumers to believe that the opinions of the creators are representative of those held by the majority of users. This may generate a pressure to act accordingly to the (mis)perceived majority, even if this is not in line with their own private opinions. Our study thus suggest that, on a global level, this condition can favour the spread of COVID-19 infodemics. 
One may ask if such interpretation is specific to fake news or, alternatively, if it can apply also to other types of news (e.g., Mainstream Media, Science) shared on  social media. While a similar group segmentation (namely, a small number of creators and a large number of consumers) is likely to occur  regardless of the content expressed (e.g., see Fig. \ref{fig:quantile}), we expect pluralistic ignorance to emerge especially with "unpopular" behaviours, such as when individuals publicly express approval for something questionable (such as sharing blatantly false content) while privately disapproving it, and do so under the false perception of a general acceptance of that behavior. Such a misalignment between public action and private preference is unlikely to occur for less questionable behaviours (e.g., sharing Mainstream Media contents). \\ 
One of the benefit of our distinction between creators and consumers is the possibility of doing a temporal analysis and therefore to judge how user behaviour and fake news circulation relate to each other. In our analysis we found that it is impossible (at least with the time resolution available to us) to understand if fake news cause a growth in the number of highly active fake news spreaders or vice versa or, again, if they simply are the effect of a third, external event. Our results suggest that the number of consumers is probably the best indicator of fake news trend, since these two quantities show a stunning cross-correlation of 0.97. This confirms the idea that it is on consumers, those more susceptible to a possible ``illusion of majority'', that one should focus in order to control and, hopefully, reduce the incidence of fake content online. In off-line contexts, the effect of pluralistic ignorance can be alleviated by informing people about the real beliefs and expectations of others \cite{berkowitz_overview_2005}. Similar kinds of ``bottom-up" interventions could be designed to prevent users to share unreliable contents in online settings. Specifically, by unveiling the collective misunderstanding on which users behaviour is based may reduce the pressure to act as the perceived majority does not let users behave according to their (potentially disagreeing) private preferences.  Such interventions may encounter less resistance and lead to less backfiring effects (e.g., increasing polarization and resistance to opinion change) compared to debunking information ``inoculated" by external authorities or fact-checkers \cite{lewandowsky_misinformation_2012, chan_debunking_2017}.  
Our study makes several important contributions to the current literature addressing the expression of opinions in social media. We provide insights on what drives individuals to share unreliable contents in online settings, complementing a growing body of interdisciplinary work exploring the social motives and the psychology that underpins the dynamics of social media sharing \cite{ren_social_2021, pennycook_psychology_2021}. 
Misinformation  spreading can  be  particularly  dangerous in the context of the current COVID-19 pandemic as it may decrease people willingness to comply with preventive behaviours  (e.g., taking vaccines,  wearing   masks  and adopting physical distancing) \cite{van2020using}. Although more evidence is needed to assess that on-line undesirable behavior may spill-over in off-line behavior (see also \cite{valensise2021lack}), shading light on the role of social factors in the context of fake news spread may offer a novel approach for fighting misinformation and it is an attempt to respond to the current call from the research community for a better integration of the social sciences to support COVID-19 pandemic response \cite{van2020using}.

\section*{Methods}
\heading{Dataset origin and description} The datasets that we used in this come from the Covid19 Infodemics Observatory \cite{gallotti_assessing_2020, covid_observatory}.
Twitter associated with the COVID-19 pandemics (coronavirus, ncov, \#Wuhan, covid19, COVID-19, sarscov2, covid) have been automatically collected using the Twitter Filter API.

The fraction of tweets included in our filter is limited by Twitter to 1\% of the total, which instead provided us with a random sub-sample of all interactions. Nevertheless,  recall of approximately 40\% of all tweets associated with Coronavirus is estimated during these months. To reconstruct the communication network from the messages, we aggregate the information from two
the three 
types of public pairwise interactions between 
users: retweets and replies

In order to identify misinformative contents we used a database of web-domains we constructed joining together multiple publicly available data sources \cite{gallotti_assessing_2020}. This allowed us to to classify the URLs, that are included in about 20\% of tweets 
as one of seven categories: Science, Mainstream Media, Satire, Clickbait, Political, Fake/Hoax, Conspiracy/Junk Science. A fraction of about 14\% of URLs were successfully classified in this sense (about 25\% in the case of Italy).
In this analysis, we considered as ``fake'' messages associated with domains marked as Clickbait, Fake/Hoax, Conspiracy/Junk Science. We also restricted our analysis to those users classified as real people, and not bots, using the machine learning methodology described in \cite{stella2019influence}.

Lastly, our analysis is focused on three countries: Italy, the United States, and the United Kingdom. To select accounts associated with a particular country, our data infrastructure is based on the geo-coding of the accounts' textual self-declared location, which we are able to successfully map to a country of the world in the 50\% of cases. Finally in the case of the US and the UK the data spanned from January 22 and May 22, 2020; for Italy the period was instead between January 22  and December 2, 2020.

\heading{Definition of Creator and Consumer groups} The definition of the concept of creators and consumers is based on a criterion applied to each user individually. This criterion is based on the fraction of fake news produced: if among all the tweets that an account has produced 20\% or more is fake, then that user is considered to be part of the creators. On the other hand, if this fraction is between 0\% and 20\%, the account is considered to be part of the consumers.\\
In order to verify that this method is actually effective in identifying two separate groups we compared our network against a null model with randomly distributed links (see Fig. \ref{fig:density}). In this way, we were able to 
highlight 
a clear separation between creators and consumers given by the fact that the former group is much more likely to be retweeted than the latter.
Furthermore, this methodology allows carrying out a sensitivity analysis to investigate the dependence of the behaviour of these systems on the threshold to be considered a creator (that we fixed to 20\%). The result of such analysis shows that no matter what threshold we choose, this criterion is effective in separating fake news spreaders responsible for producing a large part of fake news, and those responsible for retweeting them, as shown in Fig. \ref{fig:density}. Finally in the Supplementary Information we show that the creators-consumers structure is not unique to the data from the US but can be found in the data from Italy and the UK as well.

\heading{Causality detection via Convergent Cross Mapping} The algorithm that we use to infer the causal relation between the two time series in Fig. \ref{fig:causation} is called Convergent Cross Mapping. Given two time series it makes use of Taken's theorem to reconstruct the dynamical attractor that produces the first of the two and from that uses it to estimate the other. The cross correlation between the second time series and its reconstructed version is the indicator of how strongly the second is caused by the first. More information on the exact algorithm that we used can be found at \cite{rEDM}.

\heading{ACKNOWLEDGMENTS}The research reported in this work was partially supported by the EU H2020 ICT48 project ``Humane AI Net" under contract \#952026.


%

\end{document}


\begin{center}
    \textbf{SUPPLEMENTAL INFORMATION (SI)}
\end{center}

\begin{figure*}[h]
    \centering
    \includegraphics[width = 0.8\linewidth]{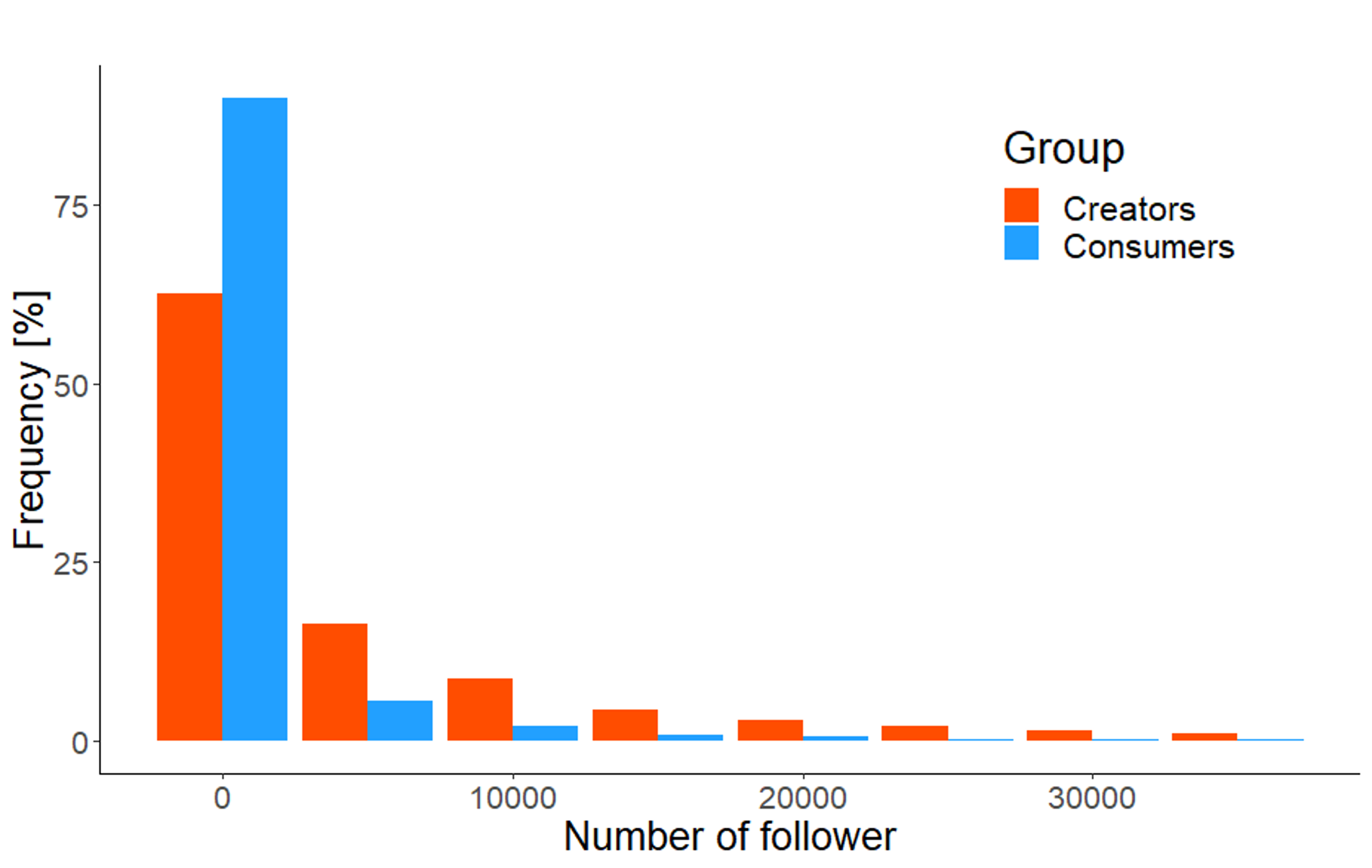}
    \caption{Distribution of the number of followers of creators and consumers. Here we included only the users in the lower 95\% percentile of the distribution, since both groups have some very strong outliers that would make the figure less readable. In particular the users with more than 40000 followers are 244 among creators and 872 among consumers. Since the number of followers of the same user changes with time we opted to use the average value. From this figure we can conclude that in the case of creators the follower distribution has an higher mean value and an heavier tail. This confirms our interpretation of creators as the main producers of fake content and responsible for its diffusion}
    \label{fig:follower_distribution}
\end{figure*}

\begin{figure*}[h]
    \centering
    \includegraphics[width = \linewidth]{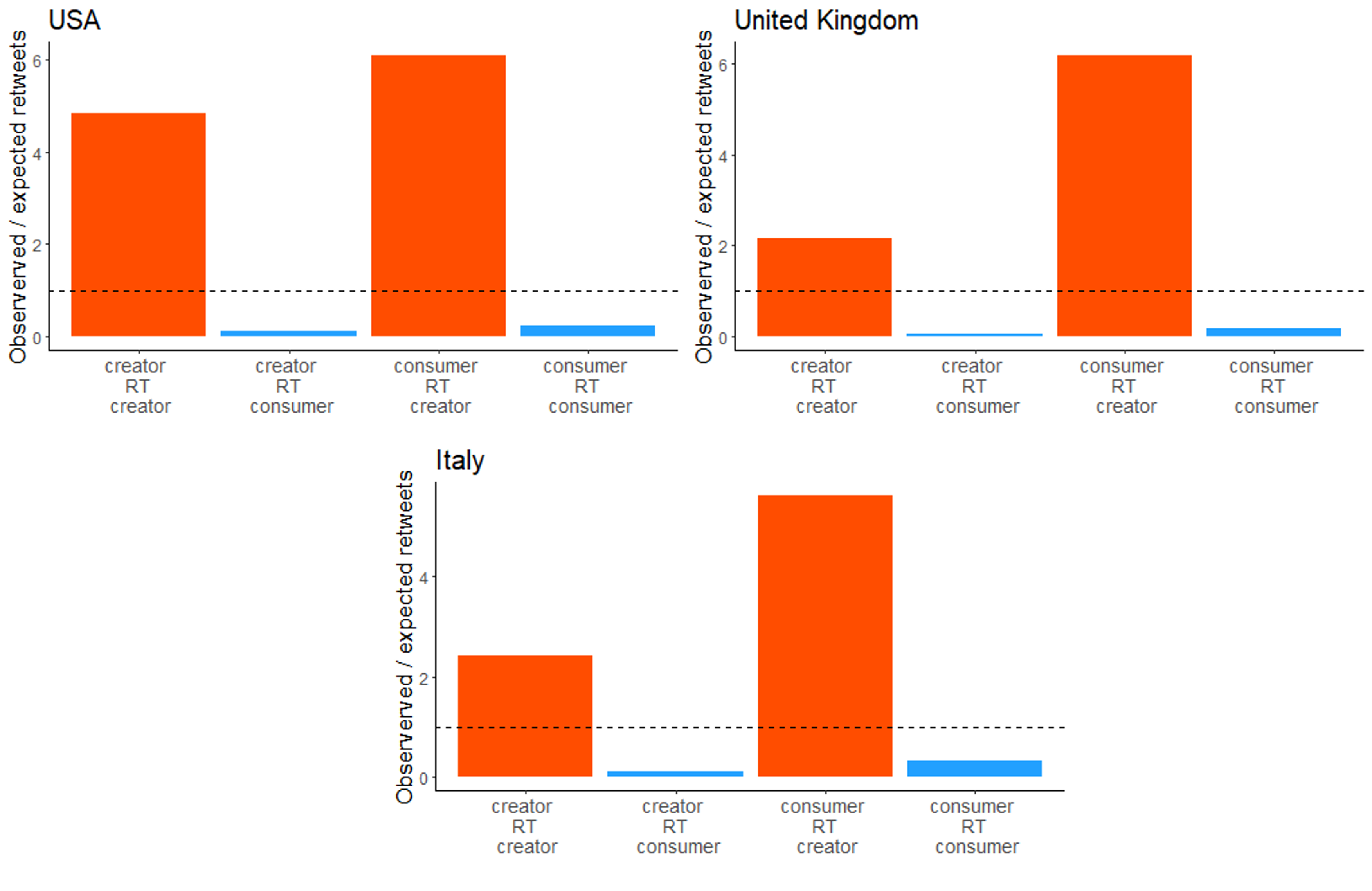}
    \caption{Comparison of the creators-consumers separation for different countries. The heights of the bars represent the ratio between the observed and expected number of links between the two groups. In this case the threshold in the percentage of most active fake news spreaders used to define the creators' group is fixed to 20\%. The behaviour seen in the US, on which we based all the analysis of our main article, is clearly visible in both Italy and the UK, with a strong density of retweets when the original tweet is coming from a creator and very low when it is coming from a consumer.}
    \label{fig:countries_comparison}
\end{figure*}

\begin{table}[h]
    \centering
    \begin{tabular}{c|ccccccccc}
        {\bfseries Percentile }& 10\% & 20\% & 30\% & 40\% & 50\% & 60\% & 70\% & 80\% & 90\% \\
        \hline
        Creators & 293.9 & 783.1 & 1513.4 & 2471.9 & 3581.8 & 5063.1 & 7667.7 & 12313.9 & 22806.3 \\
        Consumers & 49 & 115 & 208 & 342 & 543 & 895 & 1527.6 & 2752 & 5325.9 \\

    \end{tabular}
    \caption{Percentiles of the followers' distribution for creators and consumers. As in Fig. \ref{fig:follower_distribution} it can be seen that the number of followers at any percentile is lower for consumers than it is for creators. }
    \label{tab:follower-percentile}
\end{table}